\documentclass[12pt,epsfig]{article}
\usepackage{graphicx}

\begin{document}

\begin{center}
{\Large\bf Radiative Corrections to Democratic Lepton Mixing}
\end{center}

\vspace{0.3cm}
\begin{center}
{\bf Jianwei Mei} ~ and ~ {\bf Zhi-zhong Xing}
\footnote{Electronic address: xingzz@mail.ihep.ac.cn}
\end{center}
\begin{center}
{\it CCAST (World Laboratory), P.O. Box 8730, Beijing 100080, China \\
and Institute of High Energy Physics, Chinese Academy of Sciences, \\
P.O. Box 918 (4), Beijing 100049, China}
\end{center}

\vspace{2cm}
\begin{abstract}
A new ansatz of democratic lepton mixing is proposed at the GUT
scale and the radiative corrections to its phenomenological
consequences are calculated at the electroweak scale. We
demonstrate that it is possible to obtain the experimentally
favored results for both neutrino masses and lepton flavor mixing
angles from this ansatz, provided the neutrino Yukawa coupling
matrix takes a specific nontrivial pattern. The seesaw threshold
effects play a significant role in the running of relevant
physical quantities.
\end{abstract}

\newpage

To interpret the small neutrino mass-squared differences and the
large lepton flavor mixing angles observed in solar and
atmospheric neutrino oscillation experiments
\cite{SNO}--\cite{K2K}, a lot of models or ans$\rm\ddot{a}$tze of
lepton mass matrices have been proposed at low-energy scales
\cite{Review}. Among them, the scenarios based on the leptonic
flavor democracy and its explicit breaking
\cite{FX96}--\cite{Koide} are particularly simple, suggestive and
predictive. Such a phenomenologically interesting picture might
only serve as the low-scale approximation of a more fundamental
model responsible for the origin of lepton masses and flavor
mixing at a superhigh energy scale (e.g., the grand unified theory
(GUT) scale $\Lambda_{\rm GUT} \sim 10^{16}$ GeV). In this case,
the renormalization effects between high and low scales have to be
taken into account, because they are likely to modify the neutrino
mass spectrum and lepton flavor mixing parameters in a significant
way \cite{ReviewRGE}.

Possible radiative corrections to the ans$\rm\ddot{a}$tze of
democratic lepton mixing \cite{FX96}, which can naturally arise
from the slight breaking of $S(3)_L \times S(3)_R$ flavor symmetry
(i.e., flavor democracy) of the charged lepton mass matrix and
that of $S(3)$ flavor symmetry of the effective Majorana neutrino
mass matrix, have been discussed between the typical seesaw scale
($\Lambda_{\rm SS} \sim 10^{13}$ GeV) and the electroweak scale
($\Lambda_{\rm EW} \sim 10^2$ GeV) in Ref. \cite{Xing01}
\footnote{In the conventional seesaw mechanism \cite{SS} with
three heavy right-handed neutrinos $N_i$ (for $i=1,2,3$),
$\Lambda_{\rm SS}$ is sometimes referred to as $M_1$, the mass of
the lightest $SU(2)_L$ singlet $N_1$. Without loss of generality,
we take $M_3 > M_2 > M_1$ throughout this paper.}.
It is found that the mixing angle responsible for the atmospheric
neutrino oscillations, defined as $\theta_{23}$ in the standard
parametrization of the $3\times 3$ lepton flavor mixing matrix
\cite{PDG}, is rather insensitive to the renormalization effect.
Hence it is very difficult to achieve the experimentally favored
result $\theta_{23} \approx 45^\circ$ at $\Lambda_{\rm EW}$ from
the model prediction $\theta_{23} \approx 54^\circ$ \cite{FX96} at
$\Lambda_{\rm SS}$. One possible way out is to prescribe a similar
ansatz of lepton mass matrices above the seesaw scale; e.g., at or
close to the GUT scale. Then the radiative corrections from
$\Lambda_{\rm GUT}$ to $\Lambda_{\rm EW}$ will include the seesaw
threshold effects, which come from integrating out the heavy
right-handed neutrinos step by step at their mass thresholds $M_i$
(for $i=1,2,3$). Such threshold effects can drastically correct
the running behaviors of neutrino masses, flavor mixing angles and
CP-violating phases, as already shown in Ref. \cite{Threshold}.
Thus it makes sense to examine whether a constructive correction
to the democratic lepton mixing ansatz can be obtained via the
renormalization chain $\Lambda_{\rm GUT} \rightarrow M_3
\rightarrow M_2 \rightarrow M_1 \rightarrow \Lambda_{\rm EW}$.
This is just the starting point of view of this work.

We propose a new phenomenological ansatz, in which the Yukawa
coupling matrix of charged leptons results from the breaking of
flavor democracy and the effective coupling matrix of light
Majorana neutrinos is diagonal, at or close to the GUT scale. By
using the one-loop renormalization-group equations (RGEs), we
first calculate the radiative corrections to this ansatz and then
confront it with current neutrino oscillation data at low-energy
scales. To illustrate the RGE running and seesaw threshold
effects, a simple but instructive numerical example will be
presented. We demonstrate that it is possible to achieve the
experimentally favored results for neutrino masses and lepton
flavor mixing angles from our ansatz, provided the neutrino Yukawa
coupling matrix takes a specific nontrivial pattern.

\vspace{0.3cm}

At the electroweak scale, the effective Lagrangian for lepton
Yukawa interactions can be written as
\begin{equation}
-{\cal L} = \overline{E_L^{}}H_1 Y_l^{} l_R^{} - \frac{1}{2}
\overline{E_L^{}} H_2 \cdot\kappa\cdot H_2^{c\dag} E_L^c ~ + ~
{\rm h.c.} \;
\end{equation}
in the minimal supersymmetric standard model (MSSM)
\footnote{For the sake of simplicity, we assume the supersymmetry
breaking scale $\Lambda_{\rm SUSY}$ to be close to the electroweak
scale $\Lambda_{\rm EW}$. Even if $\Lambda_{\rm SUSY}/\Lambda_{\rm
EW} \sim 10$, the relevant RGE running effects between these two
scales are negligibly small for the physics under consideration
\cite{Threshold}.},
where $E_L^{}$ denotes the leptonic $SU(2)_L^{}$ doublets,
$H_1^{}$ and $H_2^{}$ are the Higgs fields, $l_R^{}$ denotes the
right-handed charged leptons, $H_2^c\equiv i \sigma_{}^2 H_2^\ast$
and $E_L^c\equiv i \sigma_{}^2 {\cal C} \overline{E_L^{}}_{}^T$
with ${\cal C}$ being the Dirac charge-conjugate matrix. After
spontaneous gauge symmetry breaking, we arrive at the charged
lepton mass matrix $M_l = vY_l\cos\beta$ and the effective
Majorana neutrino mass matrix $M_\nu = v^2\kappa \sin^2\beta$,
where $v \approx 174$ GeV and $\tan\beta$ is the ratio of the
vacuum expectation values of $H_2$ and $H_1$ in the MSSM. The
phenomenon of lepton flavor mixing, which arises from the mismatch
between the diagonalization of $Y_l$ and that of $\kappa$, is
described by the $3\times 3$ unitary matrix $V = V^\dagger_l
V_\kappa$, where
\begin{eqnarray}
&& V^\dagger_l (Y_lY^\dagger_l)V_l = \left ( \matrix{ y^2_e & 0 &
0 \cr 0 & y^2_\mu & 0 \cr 0 & 0& y^2_\tau \cr} \right ) \; ,
\nonumber \\
&& V^\dagger_\kappa \kappa V^*_\kappa = \left ( \matrix{
\kappa^{~}_1 & 0 & 0 \cr 0 & \kappa^{~}_2 & 0 \cr 0 & 0 &
\kappa^{~}_3 \cr} \right ) \; .
\end{eqnarray}
Of course, $m_\alpha = y^{~}_\alpha v\cos\beta$ (for $\alpha = e,
\mu, \tau$) and $m^{~}_i = \kappa^{~}_i v^2 \sin^2\beta$ (for
$i=1,2,3$) are the masses of charged leptons and neutrinos,
respectively. A very useful parametrization of $V$ reads
\cite{Xing04}
\begin{equation}
V = \left( \matrix{ c^{}_{12}c^{}_{13} & s^{}_{12}c^{}_{13} &
s^{}_{13} \cr -c^{}_{12}s^{}_{23}s^{}_{13} - s^{}_{12}c^{}_{23}
e^{-i\delta} & -s^{}_{12}s^{}_{23}s^{}_{13} + c^{}_{12}c^{}_{23}
e^{-i\delta} & s^{}_{23}c^{}_{13} \cr -c^{}_{12}c^{}_{23}s^{}_{13}
+ s^{}_{12}s^{}_{23} e^{-i\delta} & -s^{}_{12}c^{}_{23}s^{}_{13} -
c^{}_{12}s^{}_{23} e^{-i\delta} & c^{}_{23}c^{}_{13} } \right)
\left ( \matrix{e^{i\rho } & 0 & 0 \cr 0 & e^{i\sigma} & 0 \cr 0 &
0 & 1 \cr} \right ) \; ,
\end{equation}
where $c^{}_{ij} \equiv \cos\theta_{ij}$ and $s^{}_{ij} \equiv
\sin\theta_{ij}$ for $ij=12,23$ and $13$. Current experimental
data indicate $\theta_{12} \approx 33^\circ$, $\theta_{23} \approx
45^\circ$ and $\theta_{13} < 10^\circ$ \cite{Fit}, but there are
not any constraints on the CP-violating phases $\delta$, $\rho$
and $\sigma$. Since $\delta$ governs the strength of CP violation
in neutrino oscillations and has nothing to do with the
neutrinoless double-beta decay, it is commonly referred to as the
Dirac phase in contrast with the Majorana phases $\rho$ and
$\sigma$.

Above the seesaw scale, one encounters the neutrino Yukawa
coupling matrix $Y_\nu$ together with the Majorana mass matrix
$M_R$ of three heavy right-handed neutrinos $N_i$:
\begin{equation}
-{\cal L}^\prime = \overline{E_L^{}} H_2^{} Y_\nu^{} N +
\frac{1}{2} \overline{N^c} M_R^{} N ~ + ~ {\rm h.c.} \; .
\end{equation}
The seesaw mechanism \cite{SS} can naturally give rise to the
effective neutrino coupling matrix $\kappa = Y_\nu M^{-1}_R
Y^T_\nu$. Because three right-handed neutrinos are in general
expected to have a mass hierarchy ($M_1 < M_2 < M_3$), however, it
is necessary to take into account the seesaw threshold effects in
the RGE running chain $\Lambda_{\rm GUT} \rightarrow M_3
\rightarrow M_2 \rightarrow M_1 \rightarrow \Lambda_{\rm EW}$ step
by step (See Ref. \cite{Threshold} for a detailed description of
how to treat the RGE running through each seesaw threshold). For
simplicity, $\kappa$ and $V$ can empirically be extrapolated up to
the GUT scale, where $M_R$ can in turn be fixed by means of the
inverted seesaw formula $M_R = Y^T_\nu \kappa^{-1} Y_\nu$. In this
case, we may prescribe a phenomenological ansatz for the charged
lepton Yukawa coupling matrix $Y_l$ and the effective neutrino
coupling matrix $\kappa$ at $\Lambda_{\rm GUT}$, and calculate
radiative corrections to the lepton flavor mixing matrix $V$ at
$\Lambda_{\rm EW}$ by making use of the one-loop RGEs and by
taking account of the seesaw threshold effects.

We propose that $Y_l$ and $\kappa$ take the following forms at
$\Lambda_{\rm GUT}$:
\begin{eqnarray}
&& Y_l = \frac{c^{~}_l}{3} \left [ \left ( \matrix{
1 & 1 & 1 \cr
1 & 1 & 1 \cr
1 & 1 & 1 \cr} \right ) - \left(\matrix{%
\varepsilon^{~}_1 e^{i\phi^{~}_1} & 0 & 0 \cr 0 &
\varepsilon^{~}_2 e^{i\phi^{~}_2} & 0 \cr 0 & 0 &
\varepsilon^{~}_3 e^{i\phi^{~}_3} \cr} \right) \right ] \; ,
\nonumber \\
&& \kappa = \left ( \matrix{ \kappa^{~}_1 e^{2i\varphi^{~}_1} & 0
& 0 \cr 0 & \kappa^{~}_2 e^{2i\varphi^{~}_2} & 0 \cr 0 & 0 &
\kappa^{~}_3 e^{2i\varphi^{~}_3} \cr} \right) \; ,
\end{eqnarray}
where $\varepsilon^{~}_i$ (for $i=1,2,3$) are small perturbative
parameters; i.e., $0\leq \varepsilon_i \ll 1$. The role of
$\varepsilon^{~}_i$ is to break the flavor democracy of $Y_l$,
such that the electron and muon masses can in turn be generated.
The phase parameters $\phi^{~}_i$ of $Y_l$ and $\varphi^{~}_i$ of
$\kappa$ (for $i=1,2,3$) will contribute, respectively, to the
Dirac CP-violating phase $\delta$ and the Majorana CP-violating
phases $\rho$ and $\sigma$ of $V$ in Eq. (3). It is obvious that
the unitary matrix $V_\kappa$ defined to diagonalize $\kappa$ in
Eq. (2) is a pure phase matrix in our ansatz: $V_\kappa = {\rm
Diag} \{e^{i\varphi^{~}_1}, e^{i\varphi^{~}_2},
e^{i\varphi^{~}_3}\}$. After $Y_l$ is diagonalized by means of
$V_l$, the lepton flavor mixing matrix $V = V^\dagger_l V_\kappa$
can then be calculated.

To illustrate, let us take a simple example by assuming
$\varepsilon^{~}_1 = \varepsilon^{~}_2$, $\phi^{~}_1 = \phi^{~}_3
= 90^\circ$, $\phi^{~}_2 = 270^\circ$ and $\varphi^{~}_3 =
0^\circ$. As a result,
\begin{equation}
c^{~}_l \approx y^{~}_\tau \; , ~~~~ \varepsilon^{~}_1 =
\varepsilon^{~}_2 \approx \frac{3\sqrt{3y^{~}_e
y^{~}_\mu}}{y^{~}_\tau} \; , ~~~~ \varepsilon^{~}_3 \approx
\frac{9y^{~}_\mu}{2y^{~}_\tau} \; .
\end{equation}
In a good approximation, we obtain the lepton flavor mixing matrix
\begin{equation}
V \approx \left ( \matrix{ \displaystyle\frac{1 - a}{\sqrt{2}} &
\displaystyle\frac{1 + a}{\sqrt{2}} & \displaystyle\sqrt{2} ~ a
\cr\cr -\displaystyle\frac{1 + 3a}{\sqrt{6}} &
\displaystyle\frac{1 - 3a}{\sqrt{6}} & \displaystyle\frac{2
(1+ib)}{\sqrt{6}} \cr\cr \displaystyle\frac{1}{\sqrt{3}} &
-\displaystyle\frac{1}{\sqrt{3}} & \displaystyle\frac{1 +
ib}{\sqrt{3}} \cr} \right ) \left ( \matrix{ e^{i\varphi^{}_1 } &
0 & 0 \cr 0 & e^{i(\varphi^{}_2 + 180^\circ)} & 0 \cr 0 & 0 & 1
\cr} \right ) \; ,
\end{equation}
where $a = \sqrt{y^{~}_e/(3 y^{~}_\mu)}~$ and $b = 3
y^{~}_\mu/(2y^{~}_\tau)$. To compare between Eqs. (3) and (7), we
redefine the phases of muon and tau fields: $\mu \rightarrow \mu
e^{ib}$ and $\tau \rightarrow \tau e^{ib}$. Then the expression of
$V$ in Eq. (7) can approximately be transformed into
\begin{equation}
V \approx \left ( \matrix{ \displaystyle\frac{1 - a}{\sqrt{2}} &
\displaystyle\frac{1 + a}{\sqrt{2}} & \displaystyle\sqrt{2} ~ a
\cr\cr -\displaystyle\frac{1 + 3a}{\sqrt{6}} e^{-ib} &
\displaystyle\frac{1 - 3a}{\sqrt{6}} e^{-ib} &
\displaystyle\frac{2}{\sqrt{6}} \cr\cr
\displaystyle\frac{1}{\sqrt{3}} e^{-ib} &
-\displaystyle\frac{1}{\sqrt{3}} e^{-ib} &
\displaystyle\frac{1}{\sqrt{3}} \cr} \right ) \left ( \matrix{
e^{i\varphi^{}_1 } & 0 & 0 \cr 0 & e^{i(\varphi^{}_2 + 180^\circ)}
& 0 \cr 0 & 0 & 1 \cr} \right ) \; ,
\end{equation}
up to the accuracy of ${\cal O}(a)$ and ${\cal O}(b)$. It becomes
obvious that Eq. (8) is a reasonable simplification of Eq. (3) by
neglecting the small ${\cal O}(s^{}_{13})$ terms from its $V_{\mu
1}$, $V_{\mu 2}$, $V_{\tau 1}$ and $V_{\tau 2}$ elements. We are
therefore left with $\delta \approx b$, $\rho \approx
\varphi^{}_1$ and $\sigma \approx \varphi^{}_2 + 180^\circ$. Of
course, the values of all relevant parameters appearing in Eqs.
(6), (7) and (8) are set at $\Lambda_{\rm GUT}$.

To run the results obtained at $\Lambda_{\rm GUT}$ to the
electroweak scale by using the one-loop RGEs \cite{Threshold}, it
is necessary to fix the neutrino Yukawa coupling matrix $Y_\nu$.
Corresponding to Eq. (5), a convenient parametrization of $Y_\nu$
can be taken as
\begin{equation}
Y_\nu = y_\nu^{} U_\nu \left ( \matrix{ r^{}_1 r^{}_2 & 0 & 0 \cr
0 & r^{}_2 & 0 \cr 0 & 0 & 1 \cr} \right ) \; ,
\end{equation}
where $y^{}_\nu$, $r^{}_1$ and $r^{}_2$ are three real and
positive dimensionless parameters characterizing the eigenvalues
of $Y_\nu$, and
\begin{equation}
U_\nu = \left ( \matrix{ e^{i\xi} & 0 & 0 \cr 0 & e^{i\zeta} & 0
\cr 0 & 0 & 1 \cr} \right ) \left( \matrix{ c^{}_{1}c^{}_{3} &
s^{}_{1}c^{}_{3} & s^{}_{3} \cr -c^{}_{1}s^{}_{2}s^{}_{3} -
s^{}_{1}c^{}_{2} e^{-i\omega} & -s^{}_{1}s^{}_{2}s^{}_{3} +
c^{}_{1}c^{}_{2} e^{-i\omega} & s^{}_{2}c^{}_{3} \cr
-c^{}_{1}c^{}_{2}s^{}_{3} + s^{}_{1}s^{}_{2} e^{-i\omega} &
-s^{}_{1}c^{}_{2}s^{}_{3} - c^{}_{1}s^{}_{2} e^{-i\omega} &
c^{}_{2}c^{}_{3} } \right ) \;
\end{equation}
with $c^{}_i \equiv \cos\theta_i$ and $s^{}_i \equiv \sin\theta_i$
(for $i=1,2,3$). Because $Y_\nu$ totally involves nine free
parameters, there will be much freedom in adjusting the RGE
running behaviors of $y^{}_\alpha$ (for $\alpha = e,\mu,\tau$),
$\kappa^{}_i$ (for $i=1,2,3$) and $V$ to fit current experimental
data. This category of uncertainties is likely to be more or less
reduced in a unified model of leptons and quarks (e.g., the
$SO(10)$ model \cite{SO10}), in which the texture of $Y_\nu$ could
be related to that of quarks. Guided by the principle of
simplicity and naturalness, we shall try to pick on a reasonable
parameter space of $Y_\nu$ by avoiding possible fine-tuning of the
input parameters in our numerical calculations.

\vspace{0.3cm}

Now we present a numerical example to illustrate the RGE
corrections to the results obtained in Eqs. (6) and (8). The
eigenvalues of $Y_l$ at $\Lambda_{\rm GUT}^{}$ (i.e. $y_e^{}$,
$y_\mu^{}$ and $y_\tau^{}$) are chosen in such a way that they can
correctly run to their low-energy values \cite{PDG}. Then the
initial values of three mixing angles ($\theta^{}_{12}$,
$\theta^{}_{23}$ and $\theta^{}_{13}$) and the Dirac phase
($\delta$) of $V$ can largely be determined via Eq. (8). The mass
spectrum of three light neutrinos is assumed to have a near
degeneracy with $m^{}_1 \approx 0.245$ eV, $\Delta m^2_{21} \equiv
m^2_2 - m^2_1 >0$ and $\Delta m^2_{31} \equiv m^2_3 - m^2_1
>0$ at $\Lambda_{\rm GUT}^{}$. Thus the initial values of
$\kappa^{}_1$, $\kappa^{}_2$ and $\kappa^{}_3$ can be chosen by
using
\begin{eqnarray}
\kappa^{}_1 & = & \frac{m^{}_1}{v^2\sin^2\beta} \; ,
\nonumber \\
\kappa^{}_2 & = & \frac{\sqrt{m^2_1 + \Delta
m^2_{21}}}{v^2\sin^2\beta} \; ,
\nonumber \\
\kappa^{}_3 & = & \frac{\sqrt{m^2_1 + \Delta
m^2_{31}}}{v^2\sin^2\beta} \; ,
\end{eqnarray}
together with a typical input $\tan\beta = 10$, such that the
resultant neutrino mass-squared differences at $\Lambda_{\rm EW}$
are consistent with current solar and atmospheric neutrino
oscillation data \cite{SNO}--\cite{K2K}. Furthermore, we assume
that the eigenvalues of $Y_\nu^{}$ are strongly hierarchical
(i.e., $0 < r^{}_1 \ll 1$ and $0 < r^{}_2 \ll 1$, just like the
case of quarks) and $y_\nu^{}\sim{\cal O}(1)$ holds. It turns out
that only $\theta_1^{}$, $\theta_2^{}$, $\xi$ and $\zeta$ of
$U_\nu$ are important for the RGE running behaviors of $V$. We
find that $\theta_3^{}$ and $\omega$ of $U_\nu$ may contribute a
little to the renormalization effects on $V$, only when $r^{}_1$
and $r^{}_2$ are not so small. A summary of the input values of
relevant parameters for radiative corrections to our
phenomenological ansatz is given in Table 1, where the outputs of
$(m^{}_1, \Delta m^2_{21}, \Delta m^2_{31})$, $(\theta_{12},
\theta_{23}, \theta_{13})$ and $(\delta, \rho, \sigma)$ at
$\Lambda_{\rm EW}$ are also listed.

One can see that it is essentially possible to reproduce the
best-fit values of $\Delta m^2_{21}$, $\Delta m^2_{31}$,
$\theta_{12}$ and $\theta_{23}$, which are already determined from
a global analysis of current experimental data on solar and
atmospheric neutrino oscillations \cite{Fit}, from our lepton mass
matrices proposed at $\Lambda_{\rm GUT}$. The output $\theta_{13}
\approx 7.7^\circ$ is acceptable, because it is in no conflict
with the upper limit $\theta_{13} < 10^\circ$ set by the global
fit together with the CHOOZ experiment \cite{CHOOZ} at the $99\%$
confidence level. The output $m^{}_1 \approx 0.2$ eV implies that
the masses of three light Majorana neutrinos are nearly degenerate
at $\Lambda_{\rm EW}$, and the effective mass of the neutrinoless
double-beta decay is of the same order (i.e., $\langle
m\rangle_{ee} \approx m_1$). The latter is certainly consistent
with the present experimental upper bound $\langle m\rangle_{ee} <
0.38$ eV at the $99\%$ confidence level \cite{Fit}. We plot the
running behaviors of $m^{}_1$, $\Delta m^2_{21}$ and $\Delta
m^2_{31}$ in Fig. 1(a) and those of $\theta_{12}$, $\theta_{23}$
and $\theta_{13}$ in Fig. 1(b), from which the seesaw threshold
effects can clearly be seen. Indeed, the inverted seesaw relation
$M_R = Y^T_\nu \kappa^{-1} Y_\nu$ yields $M_1 \approx 4.1 \times
10^8$ GeV, $M_2 \approx 1.6 \times 10^{11}$ GeV and $M_3 \approx
7.2 \times 10^{13}$ GeV in our ansatz. Therefore, the most
significant radiative corrections to both neutrino masses and
lepton flavor mixing angles appear in the region between $M_3$ and
$\Lambda_{\rm GUT}$. In other words, it is the seesaw threshold
$M_3$ that can remarkably change the RGE evolution of relevant
physical parameters, such that the ansatz of democratic lepton
mixing becomes viable to fit today's low-energy neutrino data.

Finally, let us make some further remarks.

(1) It is worth mentioning that one may also discuss the
phenomenological scenario in Eq. (5) and the relevant RGE running
effects beyond the numerical example taken above. Similar results
can then be anticipated at $\Lambda_{\rm EW}$. The reason is
simply that the perturbative matrix in $Y_l$ does not affect the
dominant part of $V$, which may arise from any slight breaking of
the flavor democracy of $Y_l$.

(2) We have made use of the complicated technique developed in
Refs. \cite{ReviewRGE,Threshold} to deal with the RGE running and
seesaw threshold effects, but our phenomenological ansatz and its
low-energy consequences are new and irrelevant to the numerical
examples taken in Ref. \cite{Threshold}. The simple but typical
analysis given by us should be more suggestive and useful for
model building, in particular in the spirit of flavor democracy
and its explicit breaking at a superhigh energy scale.

(3) A number of assumptions have been taken in our treatment of
radiative corrections. Some comments on them are in order.
\begin{itemize}
\item       {\it The strong mass hierarchy of three heavy
right-handed neutrinos}. This assumption, which seems to be more
natural than the assumption of three degenerate (or partially
degenerate) right-handed neutrinos, is mainly to illustrate the
seesaw threshold effects in a more general and convincing way. Our
numerical results have shown that it is the RGE running between
$\Lambda_{\rm GUT}$ and $M_3$ that plays the dominant role in the
renormalization chain $\Lambda_{\rm GUT} \rightarrow M_3
\rightarrow M_2 \rightarrow M_1$. Such an interesting feature,
which is essentially not subject to a specific model or ansatz at
the GUT scale, has already been observed in Ref. \cite{Threshold}.

\item       {\it The strong hierarchy of three eigenvalues of
$Y_\nu$}. This assumption may become quite reasonable, if our
ansatz is embedded in the $SO(10)$ models \cite{SO10}, in which
the texture of $Y_\nu$ can be related to that of quarks with a
strong mass hierarchy. It technically simplifies our numerical
analysis, because it forbids a few free parameters of $Y_\nu$ to
play an important role in controlling the RGE evolution of $V$.

\item       {\it The approximate mass degeneracy of three light
neutrinos}. This assumption will be rather meaningful, provided
the $S(3)$ symmetry is imposed on the effective neutrino coupling
matrix $\kappa$ as a starting point of view of model building
\cite{FX96,Tanimoto}. It is also a crucial prerequisite to give
rise to significant RGE running effects, such that the ansatz
proposed at $\Lambda_{\rm GUT}$ can be compatible with current
neutrino oscillation data obtained at low energies. If the masses
of three light neutrinos are taken to be hierarchical, however,
the democratic lepton mixing ansatz under consideration will not
be viable in phenomenology.
\end{itemize}
The above discussions indicate that the parameter space considered
in our analysis is actually typical and instructive. Furthermore,
the freedom in adjusting those model parameters can be restricted
to a certain extent both by some theoretical arguments and by the
experimental data.

\vspace{0.3cm}

In summary, we have proposed a new ansatz of democratic lepton
mixing at the GUT scale and examined possible radiative
corrections to its phenomenological consequences. We show that it
is possible to obtain the experimentally favored results for both
the neutrino mass spectrum and the lepton flavor mixing angles
from this ansatz, provided the neutrino Yukawa coupling matrix
takes a specific nontrivial pattern. The seesaw threshold effects
are found to play a significant role in the RGE running of
relevant physical quantities.

Although the numerical example presented in this paper is mainly
for the purpose of illustration, it is quite suggestive for model
building. We believe that the breaking of lepton flavor democracy
at a superhigh energy scale is an interesting phenomenological
approach towards deeper understanding of the bi-large mixing
pattern of lepton flavors observed in the solar and atmospheric
neutrino oscillation experiments, and it might even hint at the
underlying flavor dynamics which governs the generation of fermion
masses and the origin of CP violation.

\vspace{0.5cm}

We would like to thank H. Fritzsch for useful discussions and
comments. This work was supported in part by the National Nature
Science Foundation of China.

\newpage

\begin{table}
\caption{A numerical example for radiative corrections to the
proposed ansatz of lepton mass matrices (from $\Lambda_{\rm GUT}$
to $\Lambda_{\rm EW}$) in the MSSM, where $\tan\beta = 10$ has
typically been taken.} \vspace{0.6cm}
\begin{center}
\begin{tabular}{|c||c|c|}\hline\hline
Parameter & Input $\left(\Lambda_{\rm GUT}^{}\right)$
& Output $( \Lambda_{\rm EW} )$ \\
\hline $m^{}_1 ({\rm eV} )$ & 0.245 & 0.2  \\
$\Delta m^2_{21} ( 10^{-5} ~{\rm eV}^2 )$ & 25 & 7.8 \\
$\Delta m^2_{31} ( 10^{-3} ~{\rm eV}^2 )$ & 9 & 2.2 \\
\hline
$\theta_{12}$ & $47.2^\circ$ & $33.6^\circ$ \\
$\theta_{23}$ & $54.4^\circ$ & $45.3^\circ$ \\
$\theta_{13}$ & $3.1^\circ$ & $7.7^\circ$ \\
$\delta$ & $5.0^\circ$ & $96.6^\circ$ \\
$\rho$ & $62.6^\circ$ & $29^\circ$ \\
$\sigma$ & $355.4^\circ$ & $302.8^\circ$ \\
\hline
$y^{}_\nu$ & 0.87 & \\
$r^{}_1$ & 0.048 & \\
$r^{}_2$ & 0.042 & \\
$\theta_1$ & $36^\circ$ & \\
$\theta_2$ & $11^\circ$ & \\
$\theta_3$ & $10^\circ$ & \\
$\omega$ & $252^\circ$ & \\
$\xi$ & $331.5^\circ$ & \\
$\zeta$ & $143^\circ$ & \\
$\varphi^{}_1$ & $62^\circ$ & \\
$\varphi^{}_2$ & $176^\circ$ & \\
\hline\hline
\end{tabular}
\end{center}
\end{table}

\begin{figure}[tbp]
\begin{center}
\includegraphics[width=11cm,height=11cm,angle=0]{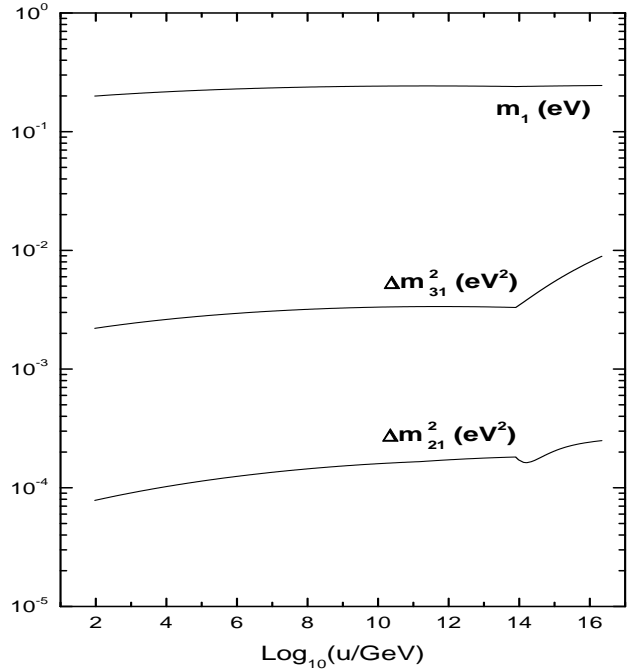}
\includegraphics[width=11cm,height=11cm,angle=0]{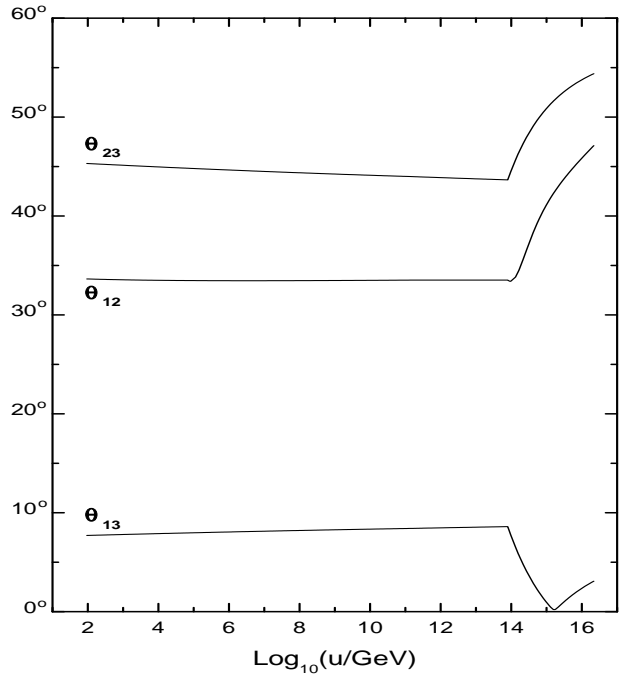}
\end{center}
\vspace{-1cm} \caption{(a) the running behaviors of $m^{}_1$,
$\Delta m^2_{21}$ and $\Delta m^2_{31}$ between $\Lambda_{\rm EW}$
and $\Lambda_{\rm GUT}$; (b) the running behaviors of
$\theta_{12}$, $\theta_{23}$ and $\theta_{13}$ between
$\Lambda_{\rm EW}$ and $\Lambda_{\rm GUT}$. The input values of
relevant parameters are listed in Table 1.}
\end{figure}

\end{document}